\documentclass[preprint,12pt]{elsarticle}
\usepackage[top=1in, bottom=1in, left=1in, right=1in]{geometry}
\usepackage{rotating}
\usepackage{graphicx}
\usepackage{amssymb}
\usepackage{lineno}
\usepackage{comment}
\usepackage{amsmath}
\usepackage{tabu}
\usepackage{xcolor}
\usepackage{multirow}
\usepackage{booktabs}
\usepackage{setspace}
\usepackage{bm}
\usepackage[colorlinks=true,bookmarksopen,pdfsubject={algorithms},linkcolor={blue}, anchorcolor={black}, citecolor={blue}, filecolor={magenta}, menucolor={black}, pagecolor={red},backref=none,urlcolor={blue}]{hyperref}
\usepackage[ruled,vlined]{algorithm2e}
\usepackage{listings}
\usepackage{xcolor}
\usepackage{tikz}
\usetikzlibrary{positioning, fit, calc}
\definecolor{codegreen}{rgb}{0,0.6,0}
\definecolor{codegray}{rgb}{0.5,0.5,0.5}
\definecolor{codepurple}{rgb}{0.58,0,0.82}
\definecolor{backcolour}{rgb}{0.95,0.95,0.92}
\usepackage{siunitx}
\usepackage{makecell}
\usepackage{geometry}

\lstdefinestyle{mystyle}{
  backgroundcolor=\color{backcolour}, commentstyle=\color{codegreen},
  keywordstyle=\color{magenta},
  numberstyle=\tiny\color{codegray},
  stringstyle=\color{codepurple},
  basicstyle=\ttfamily\footnotesize,
  breakatwhitespace=false,         
  breaklines=true,                 
  captionpos=b,                    
  keepspaces=true,                                    
  numbersep=5pt,                  
  showspaces=false,                
  showstringspaces=false,
  showtabs=false,                  
  tabsize=2
}

\journal{NPJ Computational Materials}
\begin{document}
\begin{frontmatter}

\title{Towards Computational Microscope of Chemical Order-Disorder via ML-Accelerated Monte Carlo Simulation}

\author{Fanli Zhou}
\address{School of Computer and Artificial Intelligence, Xiangnan University, Chenzhou, China}

\author{Hao Chen}
\address{Pengcheng Laboratory, Shenzhen, China}
\address{Southern University of Science and Technology, Shenzhen, China}

\author{Pengxiang Xu}
\address{Pengcheng Laboratory, Shenzhen, China}

\author{Kai Yang}
\address{Pengcheng Laboratory, Shenzhen, China}

\author{Zongrui Pei}
\address{New York University, NY, USA}

\author{Xianglin Liu\corref{cor1}$^\ast$\footnote{\footnotesize{ $^\ast$ corresponding author}}  } 
\ead{xianglinliu01@gmail.com}
\address{Pengcheng Laboratory, Shenzhen, China}

\begin{abstract}
Tailoring the performance of next-generation high entropy materials requires a deep understanding of the competition between entropy-driven random solid solution and enthalpy-driven chemical ordering. Investigating such order and disorder complexity demands atomistic simulations that achieve high accuracy, efficiency, and generalizability across vast spatial, temporal, and especially chemical scales. While machine learning (ML) interatomic potentials have transformed molecular dynamics, they remain limited in capturing diffusion-driven chemical evolution over long timescales. The recently introduced SMC-X method brings exciting opportunities. Realizing its full potential requires a comprehensive study, which is the focus of this work.
To assess model performance, we systematically benchmark invariant and equivariant architectures using a density functional theory dataset of more than 10,000 configurations spanning seven elements: Fe, Co, Ni, Al, Ti, Ta, and V.
To understand the roles of pairwise and higher-order interactions, we decouple their contributions across chemical space using an explainable machine learning approach. We also examine the impact of lattice relaxation by comparing models trained on datasets with and without structural relaxation. Our results clarify how to choose ML surrogate models for Monte Carlo simulations, bridge the gap between theory and experiment, and lay a foundation for establishing ML-accelerated Monte Carlo as a computational microscope for chemical complexity.

\end{abstract}


\end{frontmatter}

\section{Introduction}
\begin{figure} [ht!]
    \centering
    \includegraphics[width=0.9 \linewidth]{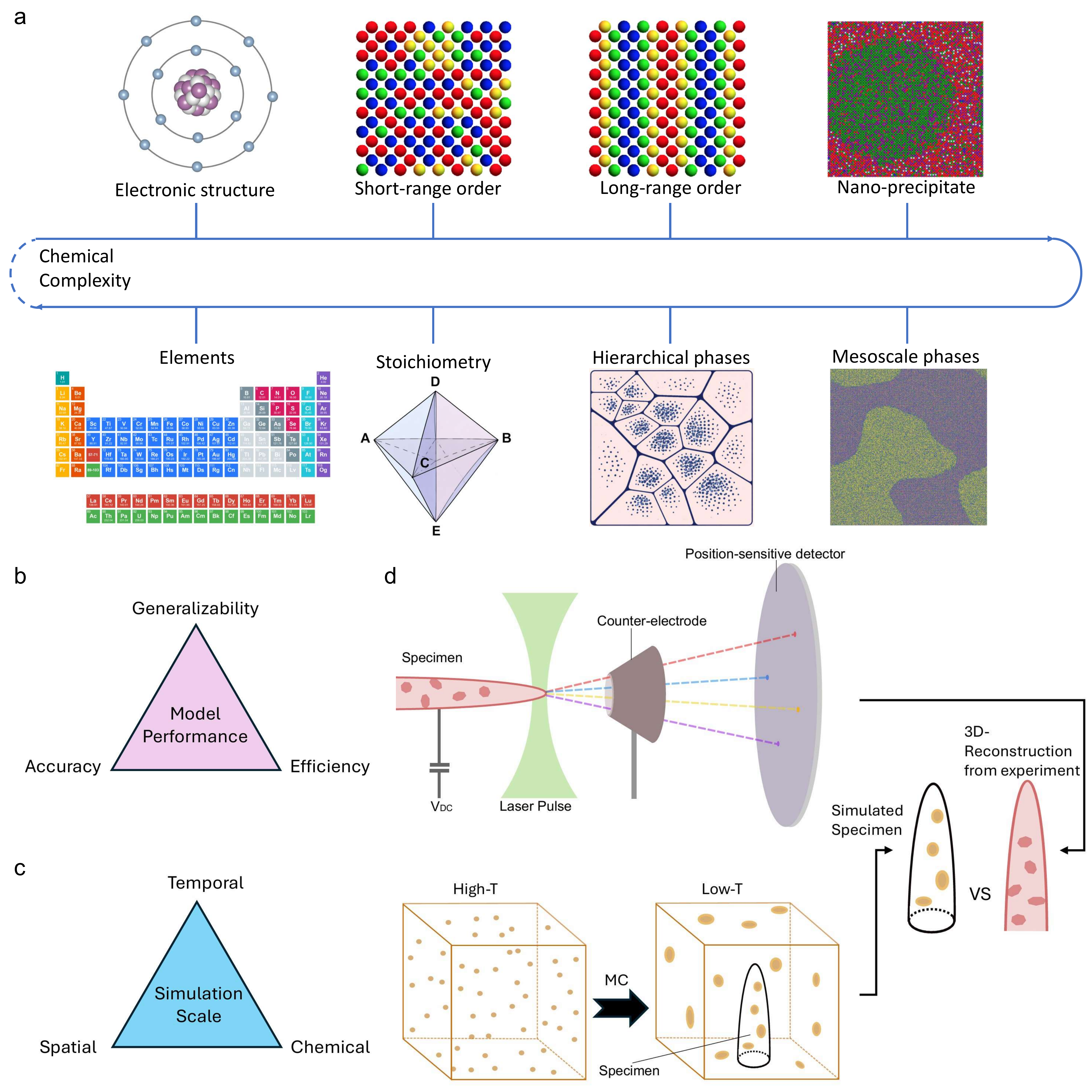}
    \caption{Schematics to illustrate the main topics of this work. (a) Various chemical complexities. (b) The AGE metrics for the performance of MLIP models. (c) The simulation scales as measured in terms of spatial, temporal, and chemical space. (d) A schematic of using ML-accelerated MC as a computational microscope for direct comparison with experiment, particularly atom probe tomography (APT).}
    \label{fig:chemical_complexity}
\end{figure}

High-entropy alloys (HEAs) \cite{George2019}, and the broader class of high-entropy materials (HEMs) \cite{Oses2020, Pacchioni2022}, represent a promising frontier in materials design, exhibiting exceptional properties across diverse applications including structural materials \cite{George2019, SENKOV2011698, WU2019444, NatureCommSai}, functional electronics \cite{Han2024, Schweidler2024}, and next-generation energy technologies \cite{doi:10.1126/sciadv.abg1600, Wang2025}. Initially, it was believed that the exceptional physical properties of HEAs mainly arise from their random solid solution structure, which is summarized by the ``four core effects'': high-entropy stabilization, sluggish diffusion, large lattice distortion, and cocktail effects \cite{yeh_2006, CoreEffectsHEA2024NRC}. However, a growing body of evidence shows that these exceptional properties often stem from an intricate competition between chemical disorder and order \cite{order-disorder-NRM-2026}. These complex patterns can span multiple length scales \cite{NanophaseHEA}, from short-range order (SRO) \cite{APT_CoCrNi_NatureMat_2024,pei2026can, SRO_MIT} and long-range order (LRO) \cite{SRO_NCS_2023} to nanoprecipitates \cite{Yang933, ORNL_HEAs_2021} and hierarchical chemical order \cite{CoNiVHEA2025Nature}, as illustrated in Fig.~\ref{fig:chemical_complexity}. The resulting chemical heterogeneity can have a profound impact on the material behavior, such as hindering dislocation motion \cite{Yin2021}, modifying phonon spectra \cite{mu_pei_liu_stocks_2018}, and creating active catalytic sites \cite{doi:10.1126/sciadv.abg1600}. This complexity, combined with the vast compositional space of HEMs, opens exciting tuning opportunities for materials design \cite{chemicalOrderNature2019, Schweidler2024, NatureComm_Tailor, YAO2024120457}.

Simulating the chemical complexity of high entropy materials requires atomistic models that simultaneously achieve high accuracy, efficiency, and generalizability, which we refer to as the AGE metrics. Here, generalizability specifically denotes a model's ability to perform reliable in-distribution interpolation and meaningful out-of-distribution extrapolation across compositional and configurational space. While high AGE performance is desirable in many areas of computational materials science, it is particularly critical for modeling order-disorder evolution in high entropy materials \cite{order-disorder-NRM-2026}. These simulations must capture essential physics across large spatial scales, from nanometers to hundreds of nanometers, extended temporal scales governed by diffusion process, and the broad chemical diversity intrinsic to high entropy alloys. Models must also maintain numerical fidelity on the order of a few meV to resolve the small energy differences arising from subtle variations in lattice site occupation. Improving AGE performance is a central goal of computational method development \cite{Lejaeghereaad3000, doi:10.1126/science.abn3445}.

The term “computational microscope” has traditionally referred to molecular dynamics \cite{computationalMicroscope, CompuitationalMicroscopeShaw}, a technique well suited for probing structural evolution and short-time, non-equilibrium processes. The rapid development of machine learning interatomic potentials (MLIPs) \cite{MPNNs_2023, doi:10.1021/jacs.4c07099} further enhances molecular dynamics, enabling physically informed models trained on high-quality first-principles datasets \cite{Horton2025, Dunn2020, UniversalModel_NCS} to achieve near ab initio accuracy while accelerating simulations by orders of magnitude \cite{doi:10.1126/science.abn3445, Kalita2025}. This combination makes MLIP-accelerated molecular dynamics a powerful tool for a wide range of applications \cite{ML_Nature2018, Nature_2021_Silicon, 2025Nature}. Despite these advances, MLIP-accelerated MD remains limited for studying order-disorder evolution, which typically occurs over seconds or longer and exceeds the timescales accessible to conventional simulations. For this purpose, Monte Carlo (MC) simulation, a complementary pillar of atomistic modeling alongside MD, is particularly effective. Instead of tracking the atomistic trajectory in real space, MC employs atomic swaps to directly sample the configuration space.  Therefore, MC describes a longer effective time scale, and has unique advantages for describing thermodynamic evolution of chemical order and disorder
\cite{Korman_npj, Yin2021, SRO_NCS_2023, liuNPJ2025}. 

Despite the rapid progress of ML-accelerated Monte Carlo (MLIP+MC) simulations in materials science, several challenges remain. First, systematic studies of their performance, particularly with respect to AGE metrics, are limited, in contrast to MLIP-accelerated MD, where benchmarking is well established \cite{Dunn2020, barrosoluque2024openmaterials2024omat24, Ong_JPC_2020, MLPAlloys2024}. Second, high-quality DFT datasets for MC are scarce, particularly compared to the millions or more configurations in MD datasets \cite{Horton2025, barrosoluque2024openmaterials2024omat24, levine2025open}. This is because, unlike MD, where a single relaxed trajectory can yield many training snapshots, MC typically requires each fully relaxed configuration as an independent training point, making data generation more expensive. Finally, conventional MC faces a fundamental scaling limitation: the sequential nature of Markov chains hinders parallelization \cite{sadigh2012scalable, PREIS20094468}. Although various methods \cite{sadigh2012scalable, mick2013gpu} have been introduced to overcome this challenge, the simulation size of MC was still limited to about a million atoms \cite{sadigh2012scalable, 2020NPJ_Li, YAO2024120457}, far below the scale routinely accessible to MD, which already reaches billion atoms \cite{10.1145/3458817.3487400}. Although several approaches have been proposed to overcome this, the recently introduced SMC-X method stands out \cite{liuNPJ2025, doi:10.1021/acs.jctc.5c01614}, demonstrating state-of-the-art performance in both throughput and system size among ML-accelerated atomistic simulation methods, and shedding light on the potential of ML-accelerated MC as a computational microscope for chemical complexity at the mesoscale. Achieving this goal requires systematic evaluation of the performance of the model, the accuracy of the simulation, and the consistency between the experiment and the theory.

The main contributions of this work are as follows. First, we systematically evaluate a range of machine learning models for Monte Carlo simulations of high entropy alloys, focusing on their accuracy, generalizability, and efficiency, as well as the roles of pairwise versus higher order interactions and invariant versus equivariant architectures. Second, we construct a high-quality dataset for a seven-element alloy system Fe, Co, Ni, Al, Ti, Ta, and V. To reduce the computational cost, we employ a linear scaling density functional theory approach, enabling efficient generation of more than 10,000 configurations. Third, using an explainable machine learning framework, we analyze how elemental contributions and interaction types affect model performance. We further examine the role of lattice relaxation and show that simple pairwise models can capture the key energetics of chemically complex alloys. Our results clarify how to choose ML surrogate models for Monte Carlo simulations, and bridge the gap between theory and experiment, therefore laying a solid foundation towards a computational microscope of chemical complexity.

\section{Results}

The configurational energy differences in high-entropy materials (HEMs) are small, typically less than a few hundred meV per atom. Accurately capturing such small energy scales in principle requires highly precise models and careful validation of the entire workflow in ML-accelerated atomistic simulations, including convergence of the DFT calculations, structural relaxation effects, and surrogate model fidelity, with accuracy within 10 meV. Achieving this level of accuracy remains a significant challenge for theoretical simulations. For example, traditional cluster expansion techniques have been shown to struggle to attain the desired energy-prediction accuracy for HEMs \cite{PhysRevB.96.014107}. In stark contrast to these theoretical challenges, simple pairwise models are widely used in practice for Monte Carlo simulations, and in many cases, show reasonable agreement with experimental observations. For instance, while it is well-known that pairwise models are not reliable for MD simulations, Ising-like models, which are effectively pairwise truncations of the cluster expansion, are frequently employed to predict the chemical evolution in high-entropy alloys (HEAs) in MC simulations \cite{Santodonato2018, MichaelWidomPair}, sometimes neglecting structural relaxation effects altogether \cite{WANG2025120635, LIU2021110135}. Several work reports that such simplified approaches can generally capture the behavior of the order-disorder transition \cite{Korman_npj, SaiNC2019, WANG2025120635, liuNPJ2025}, particularly in the relatively high temperature range \cite{Korman_npj}.
However, it remains unclear whether the apparent success of these simplified models reflects a generally valid feature of HEMs, due to, for example, error cancellation, or arises merely from special cases within specific material subclasses. To address this question, a systematic investigation of model performance across representative HEM systems is essential.

\begin{table}[htbp]
\centering
\caption{High-entropy alloys (HEAs) used in this work. Ideal configurational entropy is given in units of $R$. Valence electron concentrations (VEC) are calculated with Fe: 8, Co: 9, Ni: 10, Al: 3, Ti: 4, V: 5, Ta: 5. The atomic-size difference $\delta$ is calculated with Fe: 1.26 \AA, Co: 1.25 \AA, Ni: 1.24 \AA, Al: 1.43 \AA, Ti: 1.47 \AA, V: 1.34 \AA, Ta: 1.46 \AA. The atomic size difference is calculated by
$\delta = \sqrt{\sum_{i=1}^{n} c_i \left( 1 - \frac{r_i}{\bar{r}} \right)^2}$. ($\sigma_{yield}$: Yield Strength, $UST$: Ultimate Tensile Stress, $PSE$: Product of Strength and Elongation, $LT$: Low Temperature )}.
\label{tab:systems}
\begin{tabular}{l l c c c c c c}
\hline
Set & Alloy composition & $a$ (\AA) & $S_{\mathrm{conf}}^{\mathrm{ideal}}$ & VEC & $\delta$ & Ref. & Description\\
\hline
\multirow{7}{*}{Train}
 & Fe$_{35}$Ni$_{29}$Co$_{21}$Al$_{12}$Ta$_3$ & 3.603 & 1.414 & 8.10 & 0.052 & Nat-25 \cite{XJTU2025Nature}  & \makecell{$\sigma_{yield}=2$ GPa, \\ ductility = 25\%}\\
 & Fe$_{58}$Ni$_{33}$Al$_6$Ti$_3$            & 3.593 & 0.956 & 8.24 & 0.044 & Nat-21 \cite{ORNL_HEAs_2021} & \makecell{$UST=1.5$ GPa, \\elongation = 28\%} \\
 & Fe$_{29}$Co$_{29}$Ni$_{28}$Al$_7$Ti$_7$   & 3.551 & 1.447 & 8.22 & 0.055& Sci-19 \cite{Yang933} & \makecell{$\sigma_{yield}=1.5$ GPa, \\ductility = 50\%}\\
 & Co$_{32}$Ni$_{32}$V$_{32}$Al$_2$Ti$_2$    & 3.611 & 1.250 & 7.82 & 0.043&  Nat-25 \cite{CoNiVHEA2025Nature} & \makecell{$\sigma_{yield}^{LT}=1.2$ GPa, \\$PSE^{LT} = 76 $ GPa }\\
 & Fe$_{29}$Co$_{29}$Ni$_{28}$Al$_{14}$    & 3.619 & 1.350 & 8.15 & 0.049& NA \\
 & Fe$_{29}$Co$_{29}$Ni$_{28}$Ti$_{14}$     & 3.629 & 1.350 & 8.29 & 0.060& NA \\
 & Fe$_{34}$Co$_{33}$Ni$_{33}$              & 3.570 & 1.099 & 8.99 & 0.007& NA \\
\hline
\multirow{3}{*}{Test}
 & Fe$_{32}$Ni$_{28}$Co$_{28}$Ta$_5$Al$_7$   & 3.605 & 1.413 & 8.34 & 0.050& Nat-22 \cite{BerkeleyHEANature2022} & \makecell{low coercivity \\ high resistivity}\\
 & Fe$_{36}$Ni$_{30}$Co$_{22}$Al$_{12}$      & 3.614 & 1.317 & 8.22 & 0.046& Nat-25 \cite{XJTU2025Nature} &\makecell{$\sigma_{yield}=0.8$ GPa, \\ low ductility} \\
 & Fe$_{40}$Ni$_{33}$Co$_{24}$Ta$_3$         & 3.553 & 1.180 & 8.81 & 0.029 & Nat-25 \cite{XJTU2025Nature} &\makecell{$\sigma_{yield}=0.5$ GPa, \\ low ductility}\\
\hline
\end{tabular}
\end{table}

\subsection{Accuracy}
In this section, we focus on the accuracy of various ML models in predicting the unrelaxed DFT energies in HEAs, and the effects of structural relaxation are discussed in section~\ref{relaxation}. As summarized in Table~\ref{tab:systems}, the dataset comprises 10 face-centered cubic (FCC) high-entropy alloys (HEAs). Among them, seven have been experimentally reported to exhibit superior mechanical properties associated with nano-precipitate formation, while the remaining three systems are selected by us to increase the sampling points in compositional space. To assess generalization performance, seven alloys are used for training, and three are reserved for out-of-distribution testing. For consistency, each material includes 1,000 randomly generated configurations constructed within a 100-atom supercell. The corresponding DFT energies are computed using the MuST code with the LSMS method, without structural relaxation, as detailed in the Methods section. For analysis purpose, the ideal configurational entropy ($S_{\mathrm{conf}}^{\mathrm{ideal}}$), atomic size difference ($\delta$), and valence electron concentration (VEC) are also reported in Table~\ref{tab:systems}. The values of $S_{\mathrm{conf}}^{\mathrm{ideal}}$ range from 0.956 to 1.447, while $\delta$ spans 0.007 to 0.060. The VEC falls within a relatively narrow range for all alloys, consistent with the well-established empirical criterion for stabilizing the FCC phase in HEAs.

\begin{figure} [ht!]
    \centering
    \includegraphics[width=0.9\linewidth]{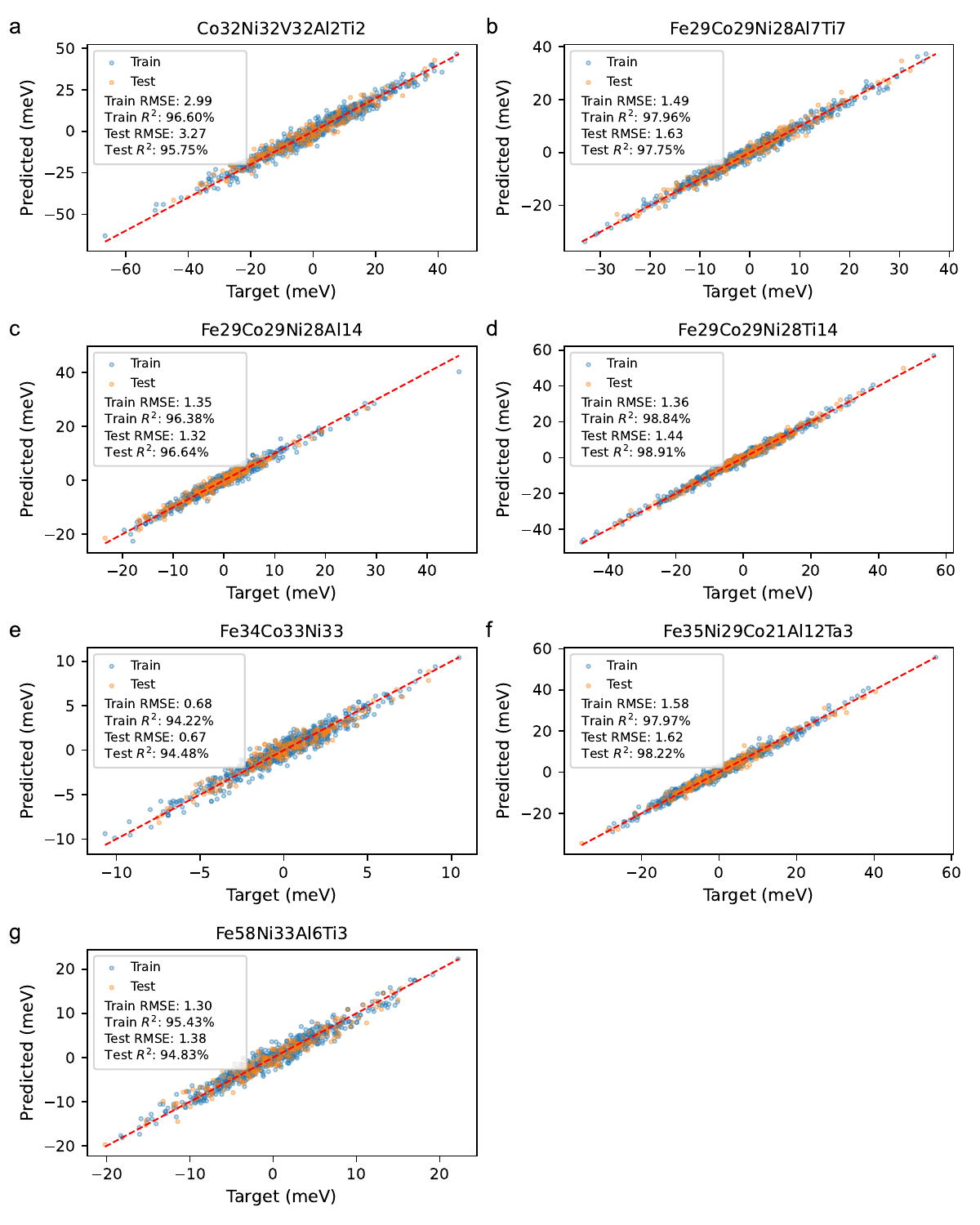}
    \caption{Parity plots of the predicted vs. DFT-calculated configurational energies for seven HEA systems using the baseline model EPI-BRR. The solid diagonal red line represents perfect agreement ($\rm{y=x}$).}
    \label{fig:parity}
\end{figure}

\begin{figure} [ht!]
    \centering
    \includegraphics[width=1\linewidth]{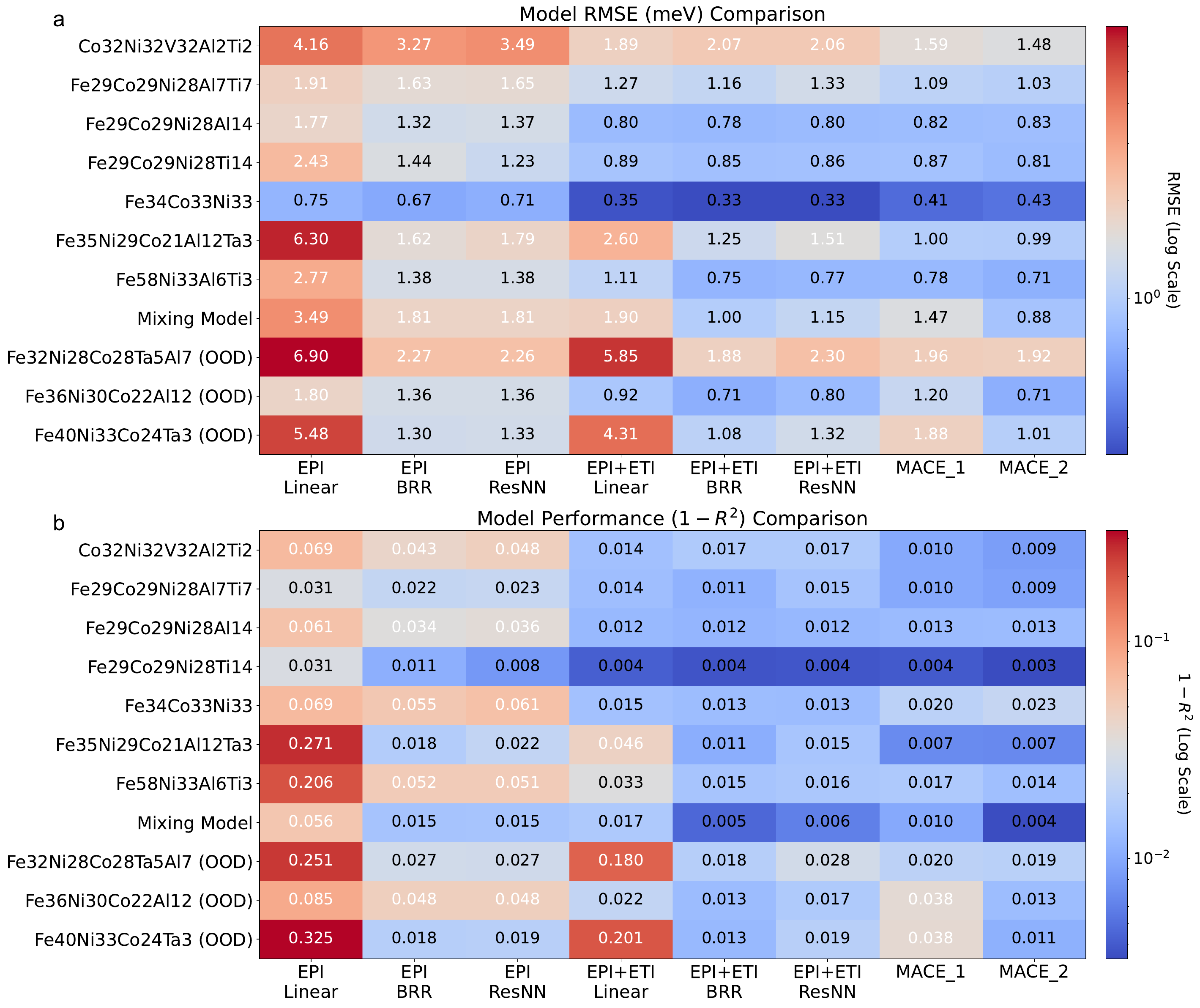}
    \caption{Heat maps of RMSE (a) and $\rm{1-R^2}$ (b) for eight different models, over different datasets. Mixing refers to combining all seven materials for training, evaluation, and testing. OOD refers to out-of-distribution test, which uses the model trained over the seven materials to test on new materials with unseen chemical compositions.}
    \label{fig:heat}
\end{figure}

We adopt Bayesian ridge regression combined with the EPI model (EPI–BRR) as the baseline. The parity plots for the seven training systems are presented in Fig.~\ref{fig:parity}. As shown, the EPI–BRR approach achieves consistently high accuracy across all materials, with test $R^2$ values ranging from 0.948 to 0.989 and root-mean-square errors (RMSEs) below 3.3 meV. These results suggest that pairwise models are often sufficient to capture the configurational energetics of high-entropy alloys (HEAs), at least for unrelaxed DFT data, consistent with previous studies.
In addition to Bayesian ridge regression, we evaluated several alternative models, with model details given in the Supplementary. Specifically, EPI-linear denotes a linear regression model optimized via stochastic gradient descent (SGD), while EPI+ETI represents using both pairwise and three-body interactions as the input features. ResNN represents models based on neural networks with residual connections. Other than invariant descriptor based models, we also include two MACE models \cite{batatia2023mace} trained with different hyperparameters, i.e., MACE\_1 with $\rm{max}_L=1$ and MACE\_2 with $\rm{max}_L=2$.
The corresponding results are shown in Fig.~\ref{fig:heat}. Notably, except for Bayesian ridge regression, all other models were trained using SGD-based optimization.

The comparative analysis (Fig.~\ref{fig:heat}) leads to the following observations: (1) when only pairwise interactions are considered, Bayesian ridge regression provides a robust and effective baseline; (2) nonlinear models, such as EPI-ResNN, do not exhibit clear performance gains over EPI–BRR; (3) training EPI models with SGD can be highly unstable, occasionally resulting in anomalously large errors; (4) incorporating three-body interactions systematically improves predictive accuracy; and (5) MACE\_2 achieves the lowest average error (0.878 meV), although its advantage over EPI+ETI with Bayesian ridge regression (1.00 meV) remains modest. (6) Pairwise models are generally more reliable for materials that more closely resemble an “ideal” HEA, namely, those composed of chemically similar elements that satisfy the Hume–Rothery rules and exhibit high configurational entropy. For example, Co$_{32}$Ni$_{32}$V$_{32}$Al$_2$Ti$_2$ shows the largest error for EPI-BRR in Fig.~\ref{fig:heat}. Consistently, this composition possesses relatively low configurational entropy, as well as pronounced chemical dissimilarity among its constituent elements, reflected by its low VEC (see Tab.~\ref{tab:systems}).

\subsection{Importance Analysis}

While the above analysis provides preliminary insights into the roles of pairwise and higher-order interaction features across different materials, it does not enable predictive generalization to unseen compositions and therefore cannot directly guide the design of high-entropy materials (HEMs). To address this limitation, we further examine the relationship between the importance of three-body interactions and the elemental composition of the material.

\begin{figure} [ht!]
    \centering
    \includegraphics[width=0.9\linewidth]{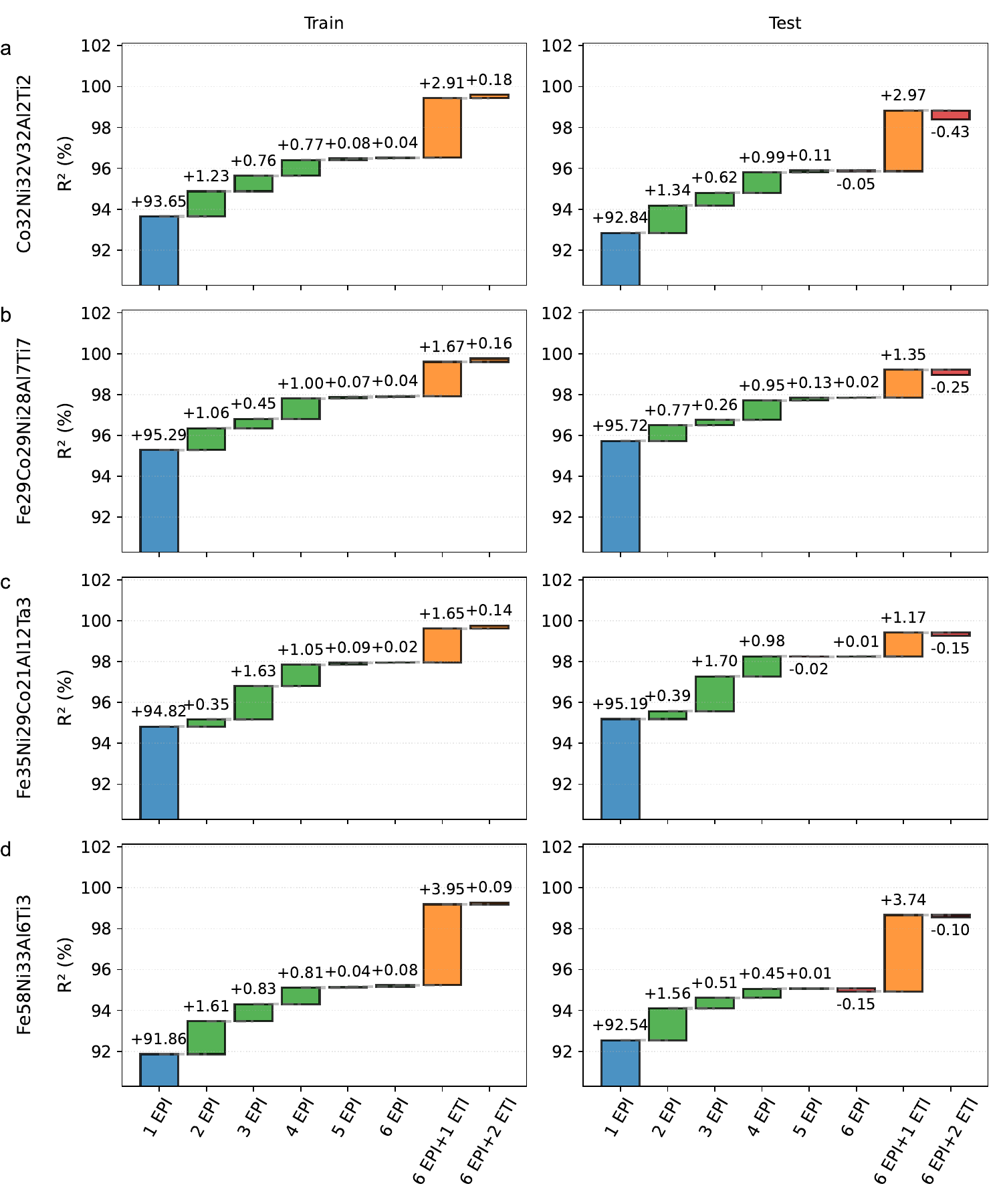}
    \caption{Hierarchical feature importance analysis of four HEA systems (a-d) using EPI and ETI BRR models. Waterfall plots illustrating the evolution of the $\rm{R^2}$ values for both training (left) and testing (right) datasets. Each bar represents the marginal increase or decrease in $\rm{R^2}$ values relative to the preceding model iteration, with the baseline established at the 1 EPI-BRR model. Bars are color-coded by feature type. }

    \label{fig:ablation}
\end{figure}

To elucidate the factors governing the predictive accuracy of the EPI and ETI models, we performed an ablation study, which is a widely used ML technique of systematically removing or modifying components of a complex model to evaluate how these changes affect its performance.
Specifically, we progressively expanded the EPI features from the first to the sixth coordination shell and quantified the marginal improvement in predictive accuracy. Subsequently, ETI features from the first to the second coordination shell were incorporated to assess the contribution of many-body correlations. The results for four representative materials are presented in Fig.~\ref{fig:ablation}.

Two main observations can be drawn. First, as the spatial range of pairwise interactions increases up to the fourth coordination shell, the model accuracy improves consistently. Beyond the fourth shell, however, the inclusion of longer-range pairwise interactions results in a slight degree of overfitting, as evidenced by the growing discrepancy between training and testing errors. Second, incorporating nearest-neighbor ETI features yields substantial improvements in predictive accuracy for all materials considered. This indicates that, although EPI provides a strong baseline representation, the inclusion of short-range many-body interactions significantly enhances model performance. In contrast, extending ETI features to longer ranges introduces mild overfitting, similar to the behavior observed for long-range pairwise interactions.

\begin{figure} [ht!]
    \centering
    \includegraphics[width=1.\linewidth]{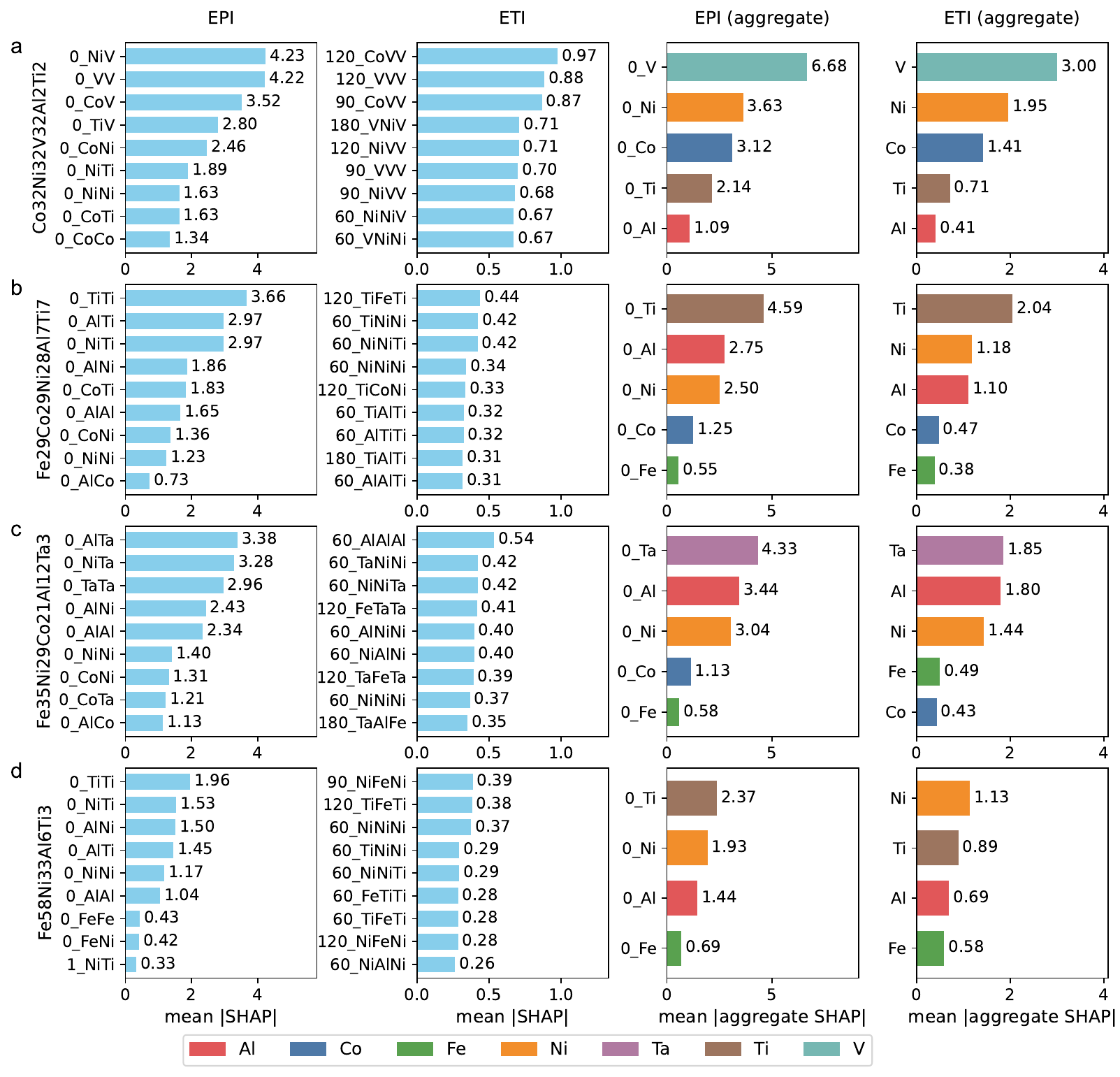}
    \caption{The Shapely value analysis of four HEA systems (a-d). The EPI descriptors are indexed by their nested shells and elements, while the ETI descriptors are defined by the angular orientation within element-specific triplets. The aggregated EPI and ETI of a specific element within a given coordination shell were calculated by integrating the SHAP values corresponding to its constituent EPI and ETI descriptors, respectively.}
    \label{fig:shapely}
\end{figure}

To facilitate material design, we further investigate the relationship between the importance of three-body interactions and elemental composition using a Shapley value analysis. The SHAP values of EPI and ETI descriptors were determined by a 6 EPI BRR model and a 6 EPI and 1 ETI BRR model, respectively. The results of four HEA systems are provided in Fig.~\ref{fig:shapely}. For Co$_{32}$Ni$_{32}$V$_{32}$Al$_2$Ti$_2$, the most influential pairwise interactions are Ni--V, V--V and Co--V. Correspondingly, the dominant higher-order contributions arise from Co--V--V and V--V--V triplets with a bond angle of $120^\circ$. In the case of Fe$_{29}$Co$_{29}$Ni$_{28}$Al$_7$Ti$_7$, the leading pairwise interactions are Ti--Ti, Al--Ti, and Ni--Ti, while the most important three-body term is the Ti--Fe--Ti triplet with a bond angle of $120^\circ$. For Fe$_{35}$Ni$_{29}$Co$_{21}$Al$_2$Ta$_3$, the top pairwise interactions are Al--Ta, Ni--Ta, and Ta--Ta, while the dominant triplet interaction is the Al--Al--Al triplet with a bond angle of $60^\circ$. In terms of Fe$_{58}$Ni$_{33}$Al$_6$Ti$_3$, the leading pairwise interactions are Ti--Ti, Ni--Ti, and Al--Ni, while the most important three-body term is the Ni--Fe--Ni triplet with a bond angle of 90. To further resolve element-specific contributions, we aggregate the Shapley values associated with each element across all interaction terms (Fig.~\ref{fig:shapely}). For Co$_{32}$Ni$_{32}$V$_{32}$Al$_2$Ti$_2$, V exhibits the largest contribution to pairwise interactions, followed by Ni. Notably, V also dominates the higher-order interaction contributions. Similarly, for Fe$_{29}$Co$_{29}$Ni$_{28}$Al$_7$Ti$_7$, Ti is the primary contributor to pairwise interactions, with Al ranking second; and Ti plays the most significant role in higher-order interactions. In terms of Fe$_{35}$Ni$_{29}$Co$_{21}$Al$_2$Ta$_3$, Ta and Al dominate both pairwise and triplet interactions. For Fe$_{58}$Ni$_{33}$Al$_6$Ti$_3$, Ti is the leading contributor to pairwise interactions, followed by Ni; however, Ni dominates three-body interactions. 

\subsection{Generalization}
As mentioned in the Introduction, a model's generalization capability can be categorized into in-distribution (ID) and out-of-distribution (OOD) performance. In this work, ID generalization refers to the model's ability to make predictions for configurations with the same chemical composition but different site occupations as those seen during training, while OOD generalization reflects its ability to predict for previously unseen chemical compositions. In the following, we will examine these two aspects of generalization separately.

For the evaluation of ID generalizability, we trained the machine learning models using the full training datasets of the seven materials in the training group and assessed their performance on the corresponding hold-out test sets. The results are summarized in the "Mixed Models" rows of Fig.~\ref{fig:heat}, with different columns representing different model architectures.
Overall, the mixed models provide reliable predictions across all seven materials, achieving RMSE values below 2~meV for all models except EPI-Linear. The inferior performance of EPI-Linear can be attributed to the large uncertainty associated with training via simple stochastic gradient descent. Consistent with the single-material models, the Bayesian Ridge approaches, including both EPI and EPI+ETI, exhibit robust and stable performance, closely followed by the ResNN models.
The MACE models demonstrate superior accuracy. In particular, MACE\_1, which effectively reduces to a pairwise model due to its small angular-momentum cutoff, already outperforms the descriptor-based EPI models. MACE\_2, incorporating explicit three-body interactions, surpasses all EPI+ETI descriptor-based models, achieving the lowest RMSE of 0.878~meV and the highest $R^2$ score of 0.996.

While ID performance reflects a model’s ability to enable reliable Monte Carlo (MC) simulations for a given material, OOD generalization is essential for deploying the model as a practical tool for materials design. To evaluate out-of-distribution capability, we used the models trained on the seven training systems to predict configuration energies for three unseen materials: Fe$_{32}$Ni$_{28}$Co$_{28}$Ta$_5$Al$_7$, Fe$_{36}$Ni$_{30}$Co$_{22}$Al$_12$, and e$_{40}$Ni$_{33}$Co$_{24}$Ta$_3$.
As shown in Fig.~\ref{fig:heat}, both MACE and EPI+ETI with BayesianRidge exhibit the strongest generalization performance. Notably, although EPI+ETI with ResNN achieves accuracy comparable to these two approaches in the ID setting, its OOD performance degrades, indicating a tendency toward overfitting. This comparison highlights the advantage of MACE in mitigating the bias–variance trade-off through the incorporation of physically motivated constraints \cite{NequIP_NC}, while EPI+ETI benefits from its relatively simple model structure and the robustness of the Bayesian framework. Finally, the EPI Bayesian Ridge model also demonstrates reliable OOD performance, although with slightly worse results compared to MACE. Specifically, the testing errors increase from 1.88, 0.71, and 1.08 meV in the ID setting to 2.27, 1.36, and 1.30 meV, respectively, for the three OOD high-entropy alloys.

\subsection{Efficiency}
Compared with accuracy and generalizability, efficiency has received relatively less attention. This is partly because, unlike accuracy and generalizability, which are intrinsic model properties, efficiency depends strongly on external factors, including computing hardware, simulation software, and how machine learning (ML) models are implemented and integrated with atomistic simulations. Nevertheless, efficiency is an essential component of the AGE metrics, as the key advantage of ML-accelerated atomistic simulations over first-principles approaches such as density functional theory (DFT) lies in their orders-of-magnitude speedup in predicting energies and forces.

\begin{figure} [ht!]
    \centering
    \includegraphics[width=0.9\linewidth]{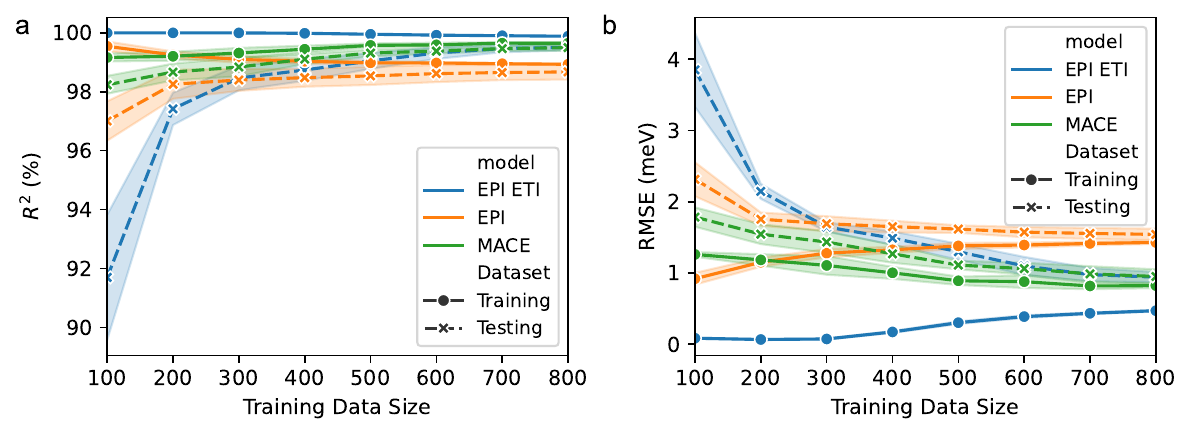}
    \caption{The learning curves for $\rm{Fe_{29}CO_{29}Ni_{28}Al_{7}Ti_{5}}$ for three different models. The $\rm{R^2}$ (a) and RMSE (b) values for the training and testing datasets of different models are plotted against training data size. The shaded areas denote the standard deviation (SD) derived from 5-fold cross-validation.}
    \label{fig:learning}
\end{figure}

Efficiency can be decomposed into training efficiency and inference efficiency. Training efficiency, which quantifies the time required to train a model, can be further divided into two components. The first is throughput, defined as the time required to complete a single epoch. The second is data efficiency, which measures the amount of data needed to reach a target level of accuracy.
Among these, data efficiency is typically the primary bottleneck, as the cost of generating training data using first-principles methods is often much higher than the cost of model training itself. While training throughput is important, it generally plays a secondary role. This is because, although equivariant models such as MACE require significantly more computational resources to train due to complex tensor operations, the real limiting factor is typically the dataset size. Since the majority of the data comes from computationally expensive first-principles calculations

To quantify the data efficiency among the most promising models identified in the previous sections, we use
 $\rm{Fe_{29}Co_{29}Ni_{28}Al_{7}Ti_{5}}$ as an example and plot the learning curves with respect to the dataset sizes, for the following models: EPI Bayesian Ridge, EPI+ETI Bayesian Ridge, and MACE. At small dataset sizes, the EPI models outperform the ETI model. However, as the dataset size increases beyond approximately 300 data points, the accuracy advantage of the EPI+ETI models becomes more apparent, which is consistent with the bias-variance tradeoff \cite{MEHTA20191}. Notably, across both small- and large-dataset ranges, the MACE model consistently outperforms both EPI and EPI+ETI Bayesian Ridge models. This further emphasizes the benefit of incorporating physical constraints, such as E(3) equivariance, into the model.

Lastly, it is important to consider the difference in the inference efficiencies. Inference efficiency is particularly critical for large-scale MC simulations, where the total number of energy evaluations is typically much higher than in molecular dynamics (MD) simulations. This is because, in MC simulations, all the local energies affected by an MC move need to be reevaluated, which often becomes the computational bottleneck. In the SMC-X framework, using a general ML energy model, the total number of energy calculations is proportional to the number of MC sweeps
$N_{sweep}$, the system size $N$, and the size of the local-interaction zone $N_{LIZ}$. 
Pairwise models, such as EPI, have a distinct advantage over non-pairwise models in terms of inference speed: they do not depend on $N_{LIZ}$ as the energy change can be computed by evaluating only the two sites directly affected by the MC swap trial. This typically results in a speedup of approximately two orders of magnitude \cite{liuNPJ2025}.

\subsection{Lattice relaxation} \label{relaxation}
At first-order approximation, high entropy alloys (HEAs) can be modeled by assuming that different elements occupy ideal, perfect lattice sites. However, this simplification overlooks the distortions in the lattice caused by the local chemical environment \cite{2024NCDistortionSNAP_HEA,pei2025rigorous}, which vary across different HEAs. For example, typical body-centered cubic (BCC) HEAs, such as MoNbTaW, tend to exhibit greater lattice distortion compared to face-centered cubic (FCC) HEAs like FeCoNiCrMn. According to the Hume-Rothery rules, the stability of solid solutions is influenced by the atomic size difference \cite{pei2020machine}, as outlined in Table ~\ref{tab:systems}. For most HEAs, except FeCoNi, the atomic size mismatch (denoted as $\delta$) is typically around 0.5, suggesting that lattice relaxation effects are non-negligible. We focus on the $\rm{Fe_{32}Ni_{28}Co_{28}Ta_{5}Al_{7}}$ and $\rm{Fe_{29}Co_{29}Ni_{28}Al_{7}Ti_{5}}$ alloys, both have superb mechanical properties originated from the formation of coherent nanoprecipitates.
\begin{figure} [ht!]
    \centering
    \includegraphics[width=1\linewidth]{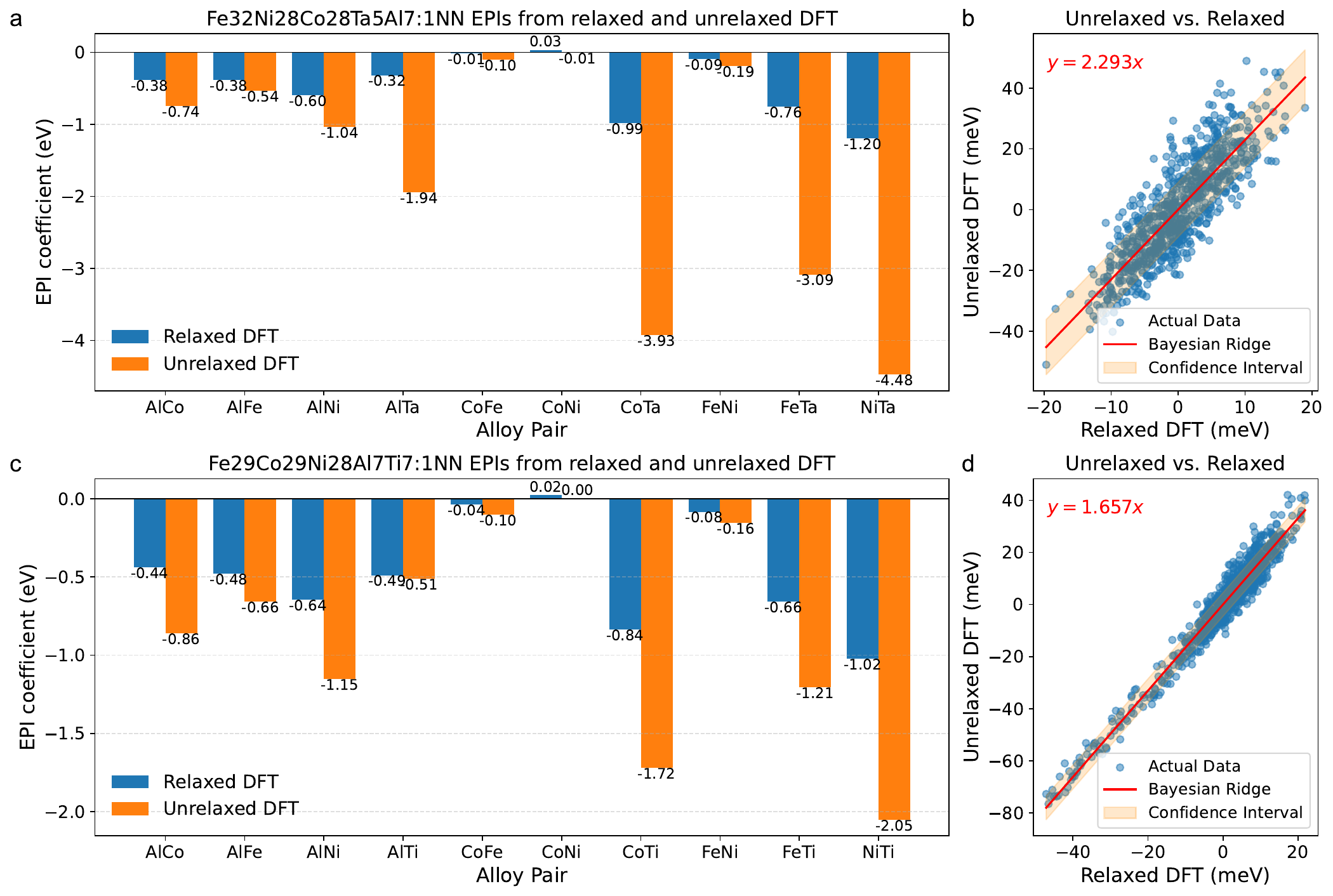}
    \caption{The first-nearest-neighbor (1NN) EPIs comparison and linear relationship between relaxed and unrelaxed data for two HEA systems, $\rm{Fe_{32}Ni_{28}Co_{28}Ta_{5}Al_{7}}$ (a, b) and $\rm{Fe_{29}CO_{29}Ni_{28}Al_{7}Ti_{5}}$ (c, d). (a, c) EPIs are extracted from a 2 EPI BRR model. (b, d) Unrelaxed data are plotted against relaxed data. The solid lines indicate the mean prediction determined by Bayesian ridge regression, with the shaded area representing the one-standard-deviation predictive uncertainty interval. }
    \label{fig:EPIs}
\end{figure}

We first compare the EPI model parameters obtained from unrelaxed calculations using the MuST code with those derived from relaxed calculations performed in QE. MuST is an all-electron method that employs approximations such as the muffin-tin potential and a local interaction cutoff, and therefore differs substantially from the plane-wave density functional theory (DFT) framework implemented in QE. As verified in the Supplementary, these methodological differences do not lead to significant discrepancies in the relative energy of different configurations, and the dominant source of variation arises from lattice relaxation. As shown in Fig.~\ref{fig:EPIs}, the EPIs extracted from relaxed DFT calculations are approximately four times smaller in magnitude than those obtained from unrelaxed configurations. This indicates that the ideal lattice configurations contain substantial residual stress, which is substantially reduced upon structural relaxation. This observation is consistent with the considerable atomic size mismatch between Al and Ta relative to FeCoNi.

It is interesting to note that, despite the pronounced difference in the magnitude of pairwise interactions, the essential features of the EPI landscape remain consistent between relaxed and unrelaxed results. For example, the lowest-energy pairs are Ni--Ta, followed by Co--Ta and Fe--Ta in both cases. In Monte Carlo simulations, this suggests that the primary effect of relaxation is to rescale the transition temperature, since the thermodynamic behavior depends on the Boltzmann factor, \( E/(k_{\mathrm{B}}T) \). Consequently, the MC results may not differ as dramatically as the large absolute energy differences might suggest, provided that the simulation temperature is appropriately rescaled.
This analysis clarifies an apparent paradox: although unrelaxed structures introduce sizable quantitative errors, they can still capture the qualitative chemical interaction trends, albeit with significantly overestimated interaction strengths. In other words, the lattice relaxation primarily affected the stress distribution within the system, but may not drastically alter the relative energy ordering for different atomic configurations. The reduction in configurational energy differences due to atomic relaxation is a general phenomenon, as observed in other systems such as Si-O~\cite{DFT_relaxation_effect_Si}, although its magnitude is system dependent. As another example, the effect is less pronounced in $\rm{Fe_{29}Co_{29}Ni_{28}Al_{7}Ti_{5}}$, as shown in the Supplementary Information. A similar observation was reported for MoNbTaW~\cite{Korman_npj}, where lattice relaxation was found to reduce the order-disorder transition temperature. 

A comparison of the relaxed and unrelaxed DFT data for the two HEAs are aslo shown Fig.~\ref{fig:EPIs} (b) and (d). It can be seen that, there is a clear linear correlation between relaxed and unrelaxed DFT results, with different slope for the two alloys, i.e. 2.293 for $\rm{Fe_{32}Ni_{28}Co_{28}Al_{7}Ta_5}$, and 1.657 for $\rm{Fe_{29}Co_{29}Ni_{28}Al_{7}Ti_{5}}$. It can also be seen that the deviation from the linear scaling relation is smaller for $\rm{Fe_{29}Co_{29}Ni_{28}Al_{7}Ti_{5}}$, as compared to $\rm{Fe_{32}Ni_{28}Co_{28}Al_{7}Ta_5}$. This result indicates that the net effect of relaxation for $\rm{Fe_{29}Co_{29}Ni_{28}Al_{7}Ti_{5}}$ is mostly a rescaling of the order-disorder transition temperature, therefore MC simulation with unrelaxed DFT data can still generate chemical profile for both the coherent precipitate phase and the matrix phase, as observed in Ref.~\cite{liuNPJ2025}. By comparison, relaxed DFT energies in $\rm{Fe_{32}Ni_{28}Co_{28}Al_{7}Ta_5}$ demonstrates much larger deviation from a simple linear scaling of the unrelaxed DFT, therefore indicating that using unrelaxed DFT for is less reliable for predicting the chemical evolution behavior. A closer look into Fig.~\ref{fig:EPIs} (a) show that the relaxation has a particularly dramatic effect on the Al-Ta pair: The EPI coefficient is reduced from the largest magnitude of 1.94 to the smallest magnitude of 0.32, as compared to all other element pairs containing Al. This substantially reduced Al-Ta interactions by relaxation seems to be the reason for the large deviation from the linear relation between relaxed and unrelaxed DFT energies in $\rm{Fe_{32}Ni_{28}Co_{28}Al_{7}Ta_5}$. To further understand the role of lattice relaxation, we also calculated the strain energy $E_s$ in an ideal fcc crystal, using the model presented in Ref.~\cite{ZongruiHR_rule}
\begin{equation}
    E_{\text{s,i}} = -\frac{\langle B \rangle}{2 \langle V \rangle} \left( \langle V_i \rangle - \langle V \rangle \right)^2,
\end{equation}
where $B$ represents the bulk modulus, $V$ represents the molar volume, and $\langle \rangle$ denotes the weighted averaging by concentration. For $\rm{Fe_{32}Ni_{28}Co_{28}Al_{7}Ta_5}$, the strain energies are calculated to be -2, -48, -43, -971, and -1653 meV/atom for Fe, Ni, Co, Al, and Ta, respectively; For $\rm{Fe_{29}Co_{29}Ni_{28}Al_{7}Ti_{5}}$, the strain energies are -4, 50, 54, 896, and 1365 meV/atom, respectively. This estimation confirms the larger energy strain in $\rm{Fe_{32}Ni_{28}Co_{28}Al_{7}Ta_5}$, which originates from its relatively large bulk modulus of 200 GPa, as compared to 110 GPa for Ti. Based on microscopic elasticity theory, a more involved model \cite{pei2025rigorous} involving Kanzaki force and lattice Green function has been developed to give a more accurate estimation of the strain energy, which is beyond the scope of this work, but interesting for future study.

\begin{figure} [ht!]
    \centering
    \includegraphics[width=1\linewidth]{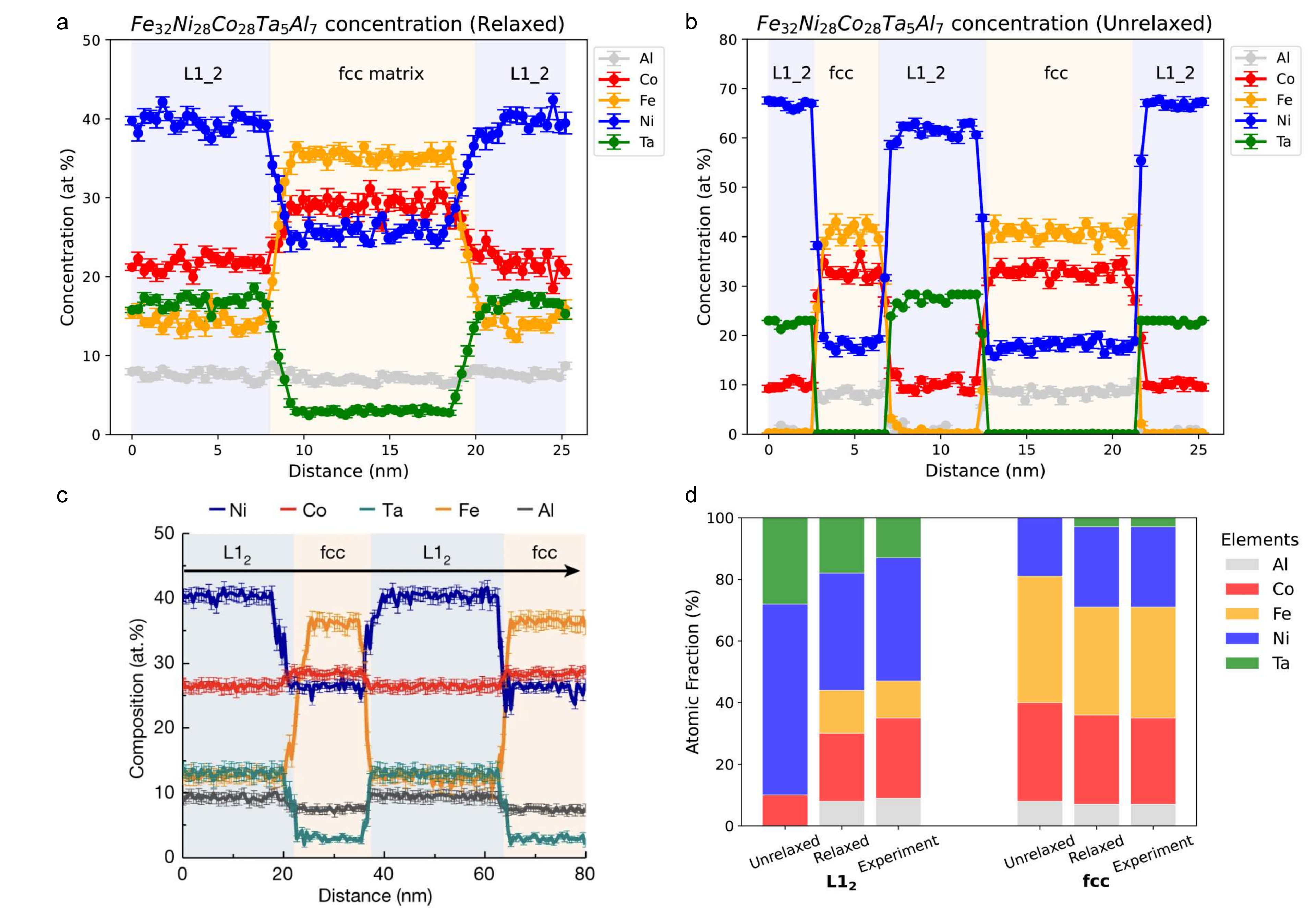}
    \caption{Comparison of the chemical composition profiles from MC simulation and experiment. (a) MC simulation results from relaxed DFT data. (b) MC simulation results from unrelaxed DFT data. (c) Experimental results from APT. Adapted from \cite{BerkeleyHEANature2022}, licensed under CC BY 4.0. (d) Bar plot comparison for precipitate and matrix phases.}
    \label{fig:Compositions}
\end{figure}

\subsection{MC as Computational Microscope}
The term "computational microscope" is traditionally synonymous with molecular dynamics, a technique uniquely suited for elucidating structural evolution and transient, short-time, non-equilibrium processes. In contrast, Monte Carlo, the other pillar of atomistic simulation, excels at deciphering chemical evolution and long-term equilibrium states. However, MC has historically been constrained by system-size limitations stemming from the sequential nature of Markov chains. With the SMC-X framework overcoming these scalability bottlenecks, it is now possible to investigate whether MC can serve as a computational microscope for chemical complexity, analogous to MD’s role in capturing structural complexity.

\begin{figure} [ht!]
    \centering
    \includegraphics[width=0.9\linewidth]{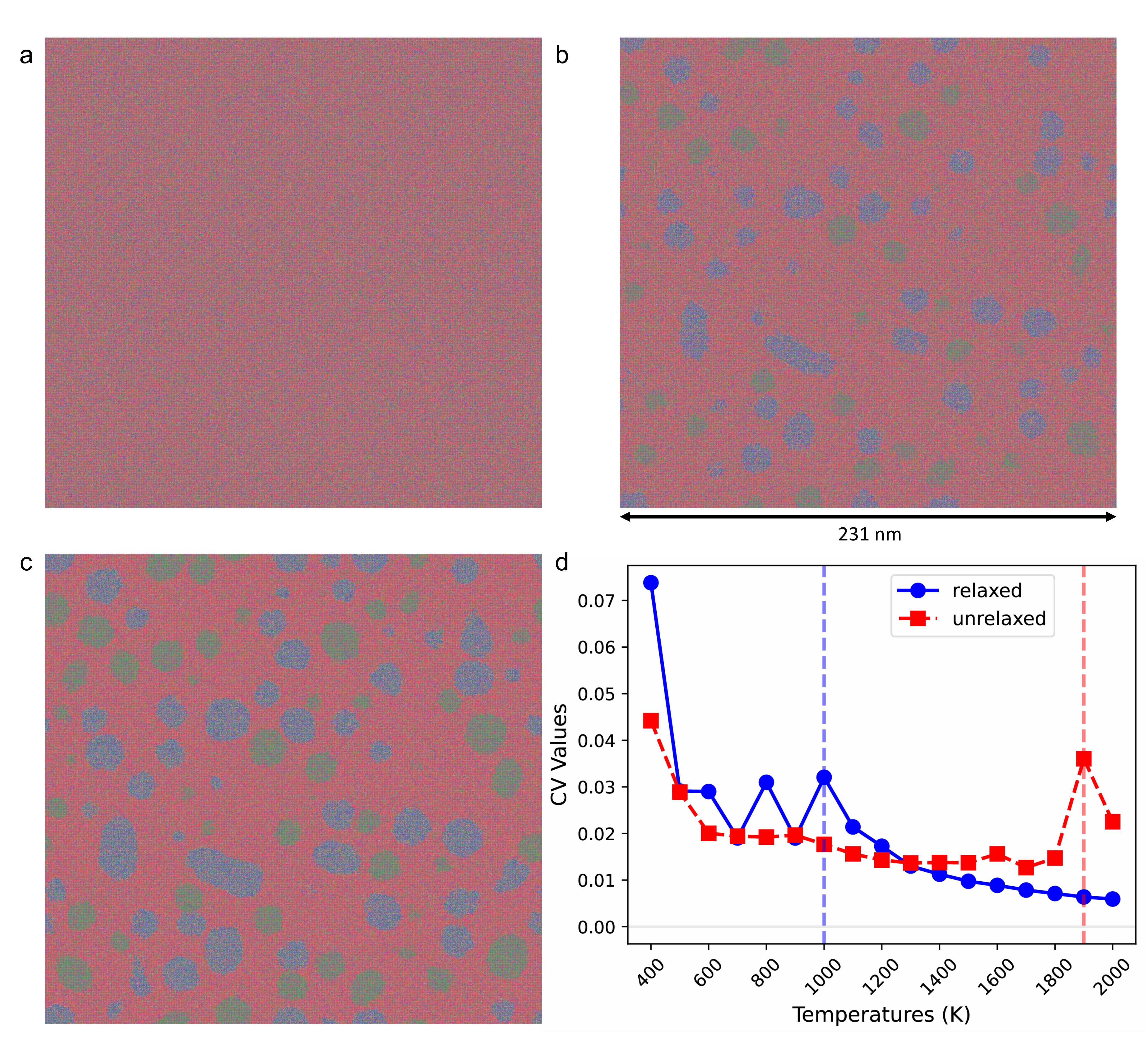}
    \caption{Configuration evolution of $\rm{Fe_{29}Co_{29}Ni_{28}Al_{7}Ti_{5}}$ at different temperatures. (a-c)The XY face of the one-billion-atom system at 1200 K, 1000K, and 800 K, respectively. (d) Specific heats from MC simulations using EPI models trained with relaxed and unrelaxed DFT data.}
    \label{fig:3Dlattice}
\end{figure}

We performed large-scale Monte Carlo simulations on a 1-million-atom $\rm{Fe_{29}Co_{29}Ni_{28}Al_{7}Ti_{5}}$ system, utilizing EPI models trained on both relaxed and unrelaxed DFT datasets, with 1 million MC steps per temperature. The resulting chemical distributions are presented in Fig.~\ref{fig:Compositions}, alongside experimental atom probe tomography data for direct comparison \cite{BerkeleyHEANature2022}. 
A comparison of Fig.~\ref{fig:Compositions} (a) and (c) reveals that the relaxed model achieves excellent agreement with experimental observations. Specifically, the Fe and Ni profiles align closely with the APT data, while Al, Co, and Ta, despite minor quantitative mismatches, accurately capture all essential structural features. In contrast, the unrelaxed model results (Fig.~\ref{fig:Compositions} (b)) exhibit significant discrepancies, most notably an exaggerated tendency for Ni to form $L1_2$ precipitates compared to the experiment. This relaxation effect is further quantified in the composition bar plots in Fig.~\ref{fig:Compositions} (d).
The thermodynamic impact is equally pronounced in the specific heat curves (Fig.~\ref{fig:3Dlattice} (d)). The relaxed model predicts an order-disorder transition temperature of approximately 1000 K, consistent with the observed stability of the alloy. Conversely, the unrelaxed model severely overestimates this transition at 1900 K. These results demonstrate that for the $\rm{Fe_{29}Co_{29}Ni_{28}Al_{7}Ti_{5}}$ system, accounting for lattice relaxation is vital for accurately predicting chemical thermo-evolution, a finding consistent with the EPI parameters discussed in the preceding section.
 
Using the relaxed model, we further investigated the morphological evolution of nanoprecipitates at 800, 1000, and 1200 K; the resulting X-Y surface projections are displayed in Fig.~\ref{fig:3Dlattice} (a–c). At 1200 K, the system remains in a disordered state with no discernible long-range ordering. As the temperature decreases to 1000 K, $L1_2$ nanoprecipitates emerge with characteristic sizes of approximately 12 nm. Upon further cooling to 800 K, these precipitates coarsen, reaching diameters of roughly 20 nm. These observations are thermodynamically consistent with the specific heat ($C_V$) profile shown in Fig.~\ref{fig:3Dlattice} (d).
It should be noted that the nanoprecipitates observed experimentally are significantly larger, with diameters of approximately 92 nm. As elucidated in our previous work \cite{doi:10.1021/acs.jctc.5c01614}, this discrepancy is mainly attributed to the fact that even 3 million MC steps are insufficient for a system of one billion atoms to achieve full thermal equilibrium.

\section{Discussion}
From a data-analytical perspective, the above results demonstrate that relatively simple pairwise models can indeed achieve surprisingly high accuracy for predicting the configurational energy of HEAs. Below, we attempt to explain the mechanism behind this effectiveness.
First, HEAs generally need to satisfy the Hume–Rothery rules to form a random solid-solution matrix phase. Consequently, the constituent elements are typically chemically and structurally similar. The large compositional complexity further tends to suppress higher-order interactions, while simultaneously generating a substantial number of pairwise features that provide a reasonably accurate description of the local chemical environment in HEAs.
Second, when dissimilar elements such as Ti, Al, or V are introduced into an FeCoNi matrix, they may promote the formation of nanoprecipitates; however, their concentrations are typically low, since higher concentrations would reduce the valence electron concentration and favor the formation of polycrystalline intermetallic phases with mixed fcc and bcc structures. In this dilute limit, the system can be viewed as a low-concentration expansion in which pairwise interactions dominate, while higher-order interactions become negligible.
Finally, although unrelaxed structures substantially overestimate configurational energies due to the absence of equilibrium atomic positions, they tend to preserve the essential attractive and repulsive tendencies among elements \cite{LatticeFeatures}. Consequently, they can still provide a qualitative estimate of nanoprecipitate formation, albeit with a systematic overestimation of transition temperatures.

From the AGE metrics across different models, we observe a general tradeoff between accuracy and efficiency. However, some models show that this tradeoff can be reduced. Data-efficient architectures such as MACE achieve strong performance despite their structural complexity. In particular, incorporating physical constraints through an E(3)-equivariant network allows MACE to retain high data efficiency while maintaining predictive accuracy, particularly for out-of-distribution (OOD) evaluations. This improvement reflects its ability to overcome the general bias–variance tradeoff \cite{MEHTA20191} by embedding physical symmetries directly into the model.
For Monte Carlo (MC) simulations, model selection should be based on a balanced evaluation of AGE performance. The EPI model trained via Bayesian regression offers at least two orders of magnitude higher computational efficiency. This advantage is critical, and sometimes essential, for large-scale simulations and high-throughput screening. As a result, despite its lower accuracy and generalizability compared to more complex models such as MACE, the EPI framework remains a practical and appealing choice when computational resources are limited. The importance of pairwise models has also recently been noted in molecular systems \cite{PairwiseNNP}.

While the current framework provides a powerful tool for achieving the goal of computational microscope of chemical order-disorder, applications requiring high predictive accuracy and extended to general HEMs will necessitate the integration of Universal Machine-Learning Interatomic Potentials (MLIPs), such as MACE, into the SMC-X simulation environment. Such an integration represents a primary vector for our future developments, bridging the gap between rapid screening and ab initio-level precision. Furthermore, while Monte Carlo (MC) methods are highly effective for modeling chemical evolution, molecular dynamics (MD) remains the most suitable—if not the only—atomistic approach for capturing structural degrees of freedom, including strain effects, dynamical loading, and melting processes, as well as revealing detailed mechanisms such as interaction between chemical order and dislocation motion. Therefore, the development of a hybrid MC/MD scheme that combines SMC-X with state-of-the-art MD methodologies is of central importance. 
Such a unified approach would pave the way toward in situ computational microscopy, capable of resolving the atomic-scale dynamical evolution of complex alloys across extended spatial, temporal, and chemical scales.

In conclusion, towards the goal of harnessing Monte Carlo simulations as a computational microscope for exploring chemical complexity up to the mesoscale, we systematically evaluated the performance of various ML models, including both invariant descriptor models and equivariant models. The results identified Bayesian Ridge linear models and MACE as offering a good combination of accuracy, generalizability, and efficiency. Specifically, when efficiency is prioritized for large-scale simulations, EPI+BRR proves advantageous, while MACE models perform exceptionally well when accuracy and generalizability are emphasized.
We also found that while relaxation dramatically affects the configurational energy in the investigated HEAs, MC simulations with an unrelaxed model can still yield surprisingly good agreement with experiments. This is because the primary effect of lattice relaxation is a linear scaling of interaction strengths, which does not drastically alter the relative energy ordering. The deviation from a linear relationship depends on the system; while it is relatively small for $\rm{Fe_{29}Co_{29}Ni_{28}Al_{7}Ti_{5}}$, incorporating relaxation has a much more significant impact on $\rm{Fe_{32}Ni_{28}Co_{28}Al_{7}Ta_5}$ due to the strong strain energy between Al and Ta. Incorporating such relaxation effects into the highly efficient SMC-X framework, we successfully obtain highly accurate predictions of the chemical profile for $\rm{Fe_{32}Ni_{28}Co_{28}Al_{7}Ta_5}$, bridging the large gap between unrelaxed results and experimental data measured with atom probe tomography.

\section{Methods}
\subsection{DFT data}
The DFT dataset contains a total of 12,000 disordered configurations, consisting of 1,000 unrelaxed for each of the 10 materials, and 1000 relaxed configurations for Fe$_{29}$Co$_{29}$Ni$_{28}$Al$_7$Ti$_7$ and Fe$_{32}$Co$_{28}$Ni$_{28}$Al$_7$Ta$_5$. We use the LSMS method \cite{LSMS} to calculate the total energy of the unrelaxed alloy structures. LSMS is a linear-scaling all-electron electronic structure calculation method. For each alloy, we use a 100-atom supercell with lattice parameters listed in Table \ref{tab:systems}. We employ a spin-polarized scheme to account for the magnetic interactions in the system. The angular momentum cutoff $l_{max}$ for the electron wavefunctions is chosen as 3, and the LIZ cutoff radius is set to 12.5 Bohr. The PBE \cite{PBE}  exchange-correlation functional is used. 
The Quantum ESPRESSO package \cite{QE} is used to calculate the total energy of the relaxed alloy structures. During structural relaxation, only the atomic positions are optimized, while the lattice vectors are kept fixed. Spin polarization is taken into account to properly describe the magnetic properties of the system. We employ pseudopotentials from the SSSP library \cite{SSSP}, with a plane-wave cutoff energy of 40 Ry and a 2×2×3 k-point grid. We perform convergence tests for the cutoff energy and k-point mesh using Fe$_{29}$Co$_{29}$Ni$_{28}$Al$_7$Ti$_7$ as a representative example. Details of the tests are in the supplementary material.

\subsection{Models}

To account for many-body effects, we incorporated effective triple interactions (ETIs) into the EPI framework to develop the EPI+ETI model. To balance the predictive accuracy with the risk of overfitting, we limited the descriptor set to six EPI shells and two ETI shells. The mathematical formalisms of those frameworks are described in the Supplementary Material.
We used machine learning methods to predict the configurational energy. In the EPI/ETI framework, the configurational energy is decomposed into a linear expansion of effective cluster interactions (ECIs). Also, the EPI/ETI parameters were estimated using linear regression (Linear), Bayesian ridge regression (BRR), or a residual neural network (ResNN). In addition to the fixed-descriptor approach, we employed MACE \cite{batatia2023mace} to capture complex many-body interactions through its equivariant message-passing architecture. Details on model configuration are documented in the Supplementary Material.

\subsection{Explanable ML}
The BRR model was used to evaluate the relative importance of different EPI and ETI descriptors in determining the total configurational energy. We first conducted a hierarchical feature importance analysis to quantify the energetic contributions of each coordination shell. Subsequently, we applied SHAP (SHapley Additive exPlanations) to interpret the model, identifying the pair and triplet clusters that most significantly influence the energy landscape. Furthermore, we calculated the aggregate SHAP values for each specific element at each structural shell to study the collective impact of specific elements. Details of the feature analysis and SHAP calculations are explained in the Supplementary Material.

\subsection{Training Protocol}
For model accuracy benchmarking, the HEA training set was split into training, validation, and test sets at 64:16:20 using a stratified sampling strategy based on configurations. For model learning curve evaluation, an alternative incremental partitioning scheme combined with MC cross-validation was employed; details are provided in the Supplementary Material. Before training, all EPI/ETI descriptors were pre-processed with Robust Standardized, which scales features based on the median and interquartile range (IQR) to minimize the impact of outliers. The details of the Stochastic Gradient Descent (SGD) optimization protocol were described in the Supplementary Material.

\subsection{SMC-X}
The EPI parameters were trained using density functional theory (DFT) data from both relaxed and unrelaxed structures. Specific heat calculations were carried out on a $60 \times 60 \times 72$ fcc supercell containing approximately a billion atoms. At each temperature, $1 \times 10^6$ Monte Carlo (MC) steps were performed, followed by $2 \times 10^4$ measurement steps. The temperature was decreased from $2000\,\mathrm{K}$ to $400\,\mathrm{K}$ in increments of $100\,\mathrm{K}$.
To obtain configurational snapshots at different temperatures, a larger supercell containing a billion atoms ($640 \times 640 \times 640$) was employed. Simulated annealing was conducted from $2000\,\mathrm{K}$ down to $800\,\mathrm{K}$ with a temperature step of $200\,\mathrm{K}$. At $800$, $1000$, and $1200\,\mathrm{K}$, $3 \times 10^6$ MC steps were performed prior to measurements, while $3 \times 10^5$ MC steps were used at higher temperatures. The composition profile was obtained by averaging over 10 independent configurations, each separated by $1 \times 10^5$ MC steps. The billion-atom simulated annealing calculation was completed in approximately $6.9\,\mathrm{h}$ using 8 NVIDIA A800 GPUs.

\subsection{Computation Hardware}
All DFT calculations were performed on a computing cluster equipped with Huawei Kunpeng 920 processors. Machine learning interatomic potential training and SMC-X simulations were conducted on a workstation equipped with an NVIDIA Quadro RTX 6000 GPU (24 GB memory), as well as on a cluster featuring eight NVIDIA A800 GPUs, each with 80 GB of HBM memory.

\section*{Code availability statement}
The code supporting the findings of this study will be available upon acceptance of this manuscript.

\section*{Data availability statement}
The data supporting the findings of this study will be available upon acceptance of this manuscript.

\section*{Acknowledgements}
The work of X. Liu and F. Zhou was supported by the National Natural Science Foundation of China under Grant 12404283. The work of F. Zhou was also supported by the High
Level Talent Start-up Fund provided by Xiangnan University. The work was also supported by Guangdong S\&T Programme under Grant 2024B0101010003. 

\section*{Author contributions}
F. Z. performed the ML training, analyzed the data, and generated the figures. X.L. designed the workflow, analyzed the data, and supervised the work. F. Z. and X. L. wrote the original manuscript. H. C. calculated the DFT data. K. Yang co-ran the SMC-X simulation. P. X. and Z. P. co-analyzed the data. All authors reviewed the manuscript.

\subsection*{Corresponding authors}
Correspondence to Xianglin Liu at xianglinliu01@gmail.com or liuxl01@pcl.ac.cn.

\bibliographystyle{model1-num-names}
\bibliography{sample.bib}

\end{document}